\documentclass[11pt]{article}%

\usepackage[b5paper,top=2.5cm,left=2cm,bottom=2.5cm,right=2cm]{geometry}
\usepackage{amsmath}
\usepackage{amssymb}
\usepackage{graphicx}
\newcommand{\vect}[1]{\mathbf{#1}}

\begin{document}
\title{A Special Class of the Scalar-Tensor Gravity with Scalar-Matter Direct Coupling
}

\author{Kim Jik-su\\
\small \textit{Pyongyang Astronomical Observatory, Academy of Sciences, Pyongyang, DPR Korea}\\
\and
Kim Chol-jun\\%
\small \textit{Physics Faculty, Kim Il Sung University, DPR Korea}\\
\small email address: cj.kim@ryongnamsan.edu.kp
}

\maketitle

\begin{abstract}
A role of the scalar-matter direct coupling in the evolution of scalar-tensor gravity was studied. If the coupling functions in the generalized scalar-tensor gravity satisfy a definite equation the scalar-tensor gravity is reduced to Einstein's tensor gravity. We consider the issue in Jordan frame and then Einstein frame. It is shown that the coupling strength in the right hand side of the master equation obtained in Damour and Nordtvedt(1993) must be added by a coupling strength $\hat{\beta}$ which characterizes the scalar-matter direct coupling. If a minimum of potential $\ln C(\varphi)$ does not coincide with one of the potential $\ln A(\varphi)$ or the potential $\ln C(\varphi)$ does not have any minimum and is a monotonic function, then the attractor mechanism of scalar-tensor gravity toward pure general relativity should be modified, and the combined potential $\ln A(\varphi)+\ln C(\varphi)$ will determine the attractor mechanism.
\end{abstract}

\paragraph{keywords:}scalar-tensor gravity; scalar-matter coupling; cosmological parameter

\section{Introduction}\label{sec1}

Damour and Nordtvedt\cite{Damour1993} had found that the general massless scalar-tensor theory of gravity relaxed toward general relativity with evolution of the Universe and it had been shown in the Einstein frame. Then, the authors applied the outcome to string dilation and obtained the similar relaxation mechanism of dilation toward general relativity\cite{Damour1994a}. In Ref.~\cite{Santiago1998}, the authors generalized the method to non-flat Universe with self-interacting potential. The relaxation mechanism of the scalar-tensor gravity to general relativity explains so small current value of the coupling strength of the scalar as $\alpha_{0}=\oldstylenums{1.4\times 10^{-2}}$\cite{Bertotti2003}.

So far many works concerning dark energy and dark matter were dealing with direct coupling between the dark ingredients and through this coupling, several cosmological problems, for example, coincidence problem, cosmological scaling, and so on are considered to be resolved, but the great majority of the works is performed on the background of Einstein's tensor gravity. Therefore, the scalar field as dark energy appears as merely an ingredient of matter.

A few works, for example, \cite{Casas1992, Damour1994b, Piazza2004, Kim2017} are dedicated to studies of the scalar-matter direct coupling in the framework of the scalar-tensor gravity, which had investigated several aspects of the coupling. Since the result of Damour and Nordtvedt\cite{Damour1993} is important in the study of the evolution of the scalar-tensor gravity we need to explore the appended effect of the scalar-matter direct coupling in the scalar-tensor gravity.

In this paper, we study the scalar-matter direct coupling in the generalized scalar-tensor gravity in both Jordan frame and in Einstein frame. In a special case, we find that the formal scalar-tensor gravity becomes, in essence, the general relativity.

In general, in Einstein frame, the attractor mechanism found by Damour and Nordtvedt\cite{Damour1993} should be modified due to the addition of scalar-matter direct coupling and a new minimum of the combined potential $\ln A(\varphi)+\ln C(\varphi)$ will control the attractor mechanism of scalar-tensor gravity toward Einstein's tensor gravity.

In Sect.~\ref{sec2}, we consider the problem in Jordan frame and find a special class of the scalar-tensor gravity. In Sect.~\ref{sec3}, the issue is considered in Einstein frame and we find two subclasses depending on the sign of the scalar-matter direct coupling strength $\beta$. In Sect.~\ref{sec4}, we give a conclusion on the obtained results.

\section{Version of Jordan frame}\label{sec2}

Ref.~\cite{Kim2017} obtained the dynamical equation for scalar field of gravity and it reads

\begin{align}\label{eq1}
2\omega \left( \phi\right)\square\phi  = &\left( {\frac{{d\ln F}}{{d\phi }} - 2\frac{{d\ln C}}{{d\phi }}} \right){T^{\left( 0 \right)}} - \frac{{d\ln F}}{{d\phi }}\left[ {\omega \left( \phi  \right){{\left( {{\phi _{,\alpha }}} \right)}^2} + \left( {4U\left( \phi  \right) + 3 \square F\left( \phi  \right)} \right)} \right] \nonumber\\
&- \frac{{d\omega \left( \phi  \right)}}{{d\phi }}{\left( {{\phi _{,\alpha }}} \right)^2} + 2\frac{{dU\left( \phi  \right)}}{{d\phi }},
\end{align}
where $F\left( \phi  \right)$, $\omega\left( \phi  \right)$ and $C\left( \phi  \right)$ are coupling functions representative for the generalized scalar-tensor gravity and $U\left( \phi  \right)$ is self-interacting potential function, which can be seen in the following action functional
\begin{equation}\label{eq2}
S = \int {{d^4}x\sqrt { - g} \left[ {\frac{1}{2}\left( {F\left( \phi  \right)R - \omega \left( \phi  \right){{\left( {{\phi _{,\alpha }}} \right)}^2}} \right) - U\left( \phi  \right) + C\left( \phi  \right)L_m^{\left( 0 \right)}\left( {\Psi ;{g_{\mu \nu }}} \right)} \right]},
\end{equation}
where we explicitly separated the scalar-matter coupling function $C(\phi)$ from the Lagrangian of matter. When $F\left( \phi  \right) = \phi ,\,\omega \left( \phi  \right) = \frac{\omega }{\phi },U\left( \phi  \right) = 0$ and $C(\phi)=1$, one recovers the simplest scalar tensor gravity, Jordan-Brans-Dicke theory. In Eq.(\ref{eq1}), $T^{\left( 0 \right)}$ is trace of $T^{(0)}{_\mu ^\nu }$ and the latter is energy-momentum tensor derived from $L_m^{(0)}$.(In Ref.~\cite{Kim2017}, $\omega(\phi)$ is set one)

The issue we must focus our attention is the following. If one assumes the first term of r.h.s. in Eq.~\eqref{eq1} to vanish
\begin{align}\label{eq3}
\left( {\frac{{d\ln F}}{{d\phi }} - 2\frac{{d\ln C}}{{d\phi }}} \right)=0,
\end{align}
then Eq.~\eqref{eq1} becomes
\begin{align}\label{eq4}
2\omega \left( \phi\right)\square\phi  = - \frac{{d\ln F}}{{d\phi }}\left[ {\omega \left( \phi  \right){{\left( {{\phi _{,\alpha }}} \right)}^2} + \left( {4U\left( \phi  \right) + 3 \square F\left( \phi  \right)} \right)} \right]- \frac{{d\omega \left( \phi  \right)}}{{d\phi }}{\left( {{\phi _{,\alpha }}} \right)^2} + 2\frac{{dU\left( \phi  \right)}}{{d\phi}}.
\end{align}

Therefore the indirect coupling of scalar to matter $T^{(0)}$ disappears and the scalar field ceases to be a component of the gravitational field. In Jordan-Brans-Dicke gravity, Eq.~\eqref{eq4} becomes
\begin{align}\label{eq5}
\left(2\omega+3\right)\square\phi=0.
\end{align}

In Jordan-Brans-Dicke gravity, if one puts $C(\phi\phi^n$, where $n$ is a constant, we get
\begin{align}\label{eq6}
\left(2\omega+3\right)\square\phi=(1-2n)T^{(0)}.
\end{align}

If $n=\frac{1}{2}$ holds, the gravitation ceases to be scalar-tensorial and the scalar becomes a pure matter field.

Ref.~\cite{Casas1992} first considered the direct coupling between scalar and matter in Jordan-Brans-Dicke gravity and obtained the same equation as Eq.~\eqref{eq6} putting $C(\phi)=\phi^n$ as above.

Ref.~\cite{Casas1992} pointed out that the case $n=\frac{1}{2}$ renders the scalar-tensor gravity to reduce to Einstein's tensor one, but authors merely declare that the scalar then is constant.

Eq.~\eqref{eq5} has trivially a constant solution $\phi=$const. But, besides, it has dynamical solutions of the scalar field. Eq.~\eqref{eq5} shows that the scalar manifests exclusively as a matter field in Einstein's tensor gravity background. However, because of the scalar-matter direct coupling, in the case of the dynamical solution of scalar, trajectories of test particles do not follow geodesics and deviate from them. Instead of geodesics, the kinetic equation of a test particle reads
\begin{align}\label{eq7}
\frac{d^2 x^\mu}{d\tau^2}+\Gamma^\mu_{\rho\sigma}\frac{d x^\rho}{d\tau}\frac{d x^\sigma}{d\tau}+(\ln C(\phi))^{,\mu}=0,
\end{align}
The last term in Eq.~\eqref{eq7} makes a particle deviate from geodesic, and it may be interpreted as ``a friction force'' as Ref.~\cite{Casas1992} dubbed.

Eq.~\eqref{eq3} is integrated to give a solution
\begin{align}\label{eq8}
F(\phi)=\text{const } C^2(\phi).
\end{align}
Here, the constant may be set one without loss of generality. We will refer to it as a special class of scalar-tensor gravity if Eq.~\eqref{eq3} or \eqref{eq8} is satisfied. Next we can consider the cases that Eq.~\eqref{eq3} or \eqref{eq8} is not satisfied.

As Eq.~\eqref{eq1} shows, the first term in right hand side which represents the indirect coupling of the scalar to matter, is changed compared with the case of vanishing $d\ln C(\phi)/d\phi$.(i.e. the case $C(\phi)=1$) Such a modification of the first term in Eq.~\eqref{eq1} due to the scalar-matter direct coupling($C(\phi)\neq1$) varies the strength of indirect coupling of scalar to matter.

More detailed consideration will be given in Einstein frame below.

\section{Version of Einstein frame}\label{sec3}

The conformal transformation from Jordan frame to Einstein one is performed by the following one of metric tensor
\begin{align}\label{eq9}
\tilde{g}_{\mu\nu}=F(\phi)g_{\mu\nu}=\frac{1}{A^2(\varphi)}g_{\mu\nu}
\end{align}
Then, the scalar field must be redefined so as to express the kinetic term in canonical form
\begin{align}\label{eq10}
\varphi=\int {d\phi\frac{1}{F(\phi)}\sqrt{\omega(\phi)F(\phi)+\frac{3}{2}\left( \frac{dF(\phi)}{d\phi}\right)^2}} 
\end{align}

The action in Einstein frame becomes
\begin{align}\label{eq11}
S = \int {{d^4}x\sqrt { - \tilde{g}} \left[ {\frac{1}{2}\left( \tilde{R}-2\tilde{g}^{\mu\nu}\varphi_{,\mu}\varphi_{,\nu} \right) - \tilde{U}\left( \phi  \right) + \tilde{C}\left( \phi  \right)\tilde{L}_m^{( 0 )}\left( {\psi, A^2(\varphi){\tilde{g}_{\mu \nu }}} \right)} \right]}.
\end{align}
The coupling function, potential function and $\tilde{L}_m^{(0)}$, then are transformed as follows:	
\begin{align}
\label{eq12}\tilde{C}(\varphi)&=C[\phi(\varphi)]\\
\label{eq13}\tilde{V}(\varphi)&=A^4(\varphi)U[\phi(\varphi)]\\
\label{eq14}\tilde{L}_m^{(0)}(\varphi)&=A^4(\varphi)L_m^{(0)}
\end{align}
and the energy-momentum tensor is transformed
\begin{subequations}
\begin{align}
\tilde{T}^{\mu\nu}&=A^6T^{\mu\nu},\\
\tilde{T}_{\mu}^{\nu}&=A^4T_{\mu}^{\nu},\\
\tilde{T}_{\mu\nu}&=A^2T_{\mu\nu}.
\end{align}
\end{subequations}

Variations of the action with respect to $\tilde{g}_{\mu\nu}$ and $\varphi$ yield the dynamical equations of gravitation.

Matter is described by a perfect-fluid representation in both conformal frames
\begin{align}
T_{\mu\nu}=(\rho+P)u_{\mu}u_{\nu}+Pg_{\mu\nu},\\
\tilde{T}_{\mu\nu}=(\tilde{\rho}+\tilde{P})\tilde{u}_{\mu}\tilde{u}_{\nu}+\tilde{P}\tilde{g}_{\mu\nu},
\end{align}
where $g_{\mu\nu}u^{\mu}u^{\nu}=\tilde{g}_{\mu\nu}\tilde{u}^{\mu}\tilde{u}^{\nu}$ and
\begin{align}
\tilde{\rho}=A^4\rho,\\
\tilde{P}=A^4P.
\end{align}

We consider the evolution of the scalar in FRW universe described by the following metric
\begin{align}
ds^2=-dt^2+a^2(t)d\vect{x}^2.
\end{align}

Instead of $a(t)$, introduce the new evolution variable as in Ref.~\cite{Damour1993}
\begin{align}\label{eq21}
p=\ln\frac{\tilde{a}}{\text{const}}.
\end{align}

Then, the master equation for the scalar field is written as follows
\begin{align}\label{eq22}
\frac{2(\tilde{\rho}+\tilde{U})}{3-(\varphi')^2}\varphi''+[(\tilde{\rho}-\tilde{P})+2\tilde{U}]\varphi'=-(\alpha+\hat{\beta})(\tilde{\rho}-3\tilde{P})-\frac{d\tilde{U}(\varphi)}{d\varphi},
\end{align}
where primes denote derivatives with respect to $p$ and
\begin{align}
\label{eq23}\alpha(\varphi)&\equiv\frac{d\ln A(\varphi)}{d\varphi},\\
\label{eq24}\hat{\beta(\varphi)}&\equiv \frac{d\ln \tilde{C}(\varphi)}{d\varphi}=\frac{d\ln C[\phi(\varphi)]}{d\varphi},\\
\label{eq25}\tilde{\rho}(\varphi)&=\tilde{C}(\varphi)\tilde{\rho}^{(0)}(\varphi),\\
\label{eq26}\tilde{P}(\varphi)&=\tilde{C}(\varphi)\tilde{P}^{(0)}(\varphi).
\end{align}
Here, $\tilde{\rho}^{(0)}(\varphi)$ and $\tilde{P}^{(0)}(\varphi)$ are quantities derived from $\tilde{L}^{(0)}_m(\varphi)$, (Eq.~\ref{eq14}) $\tilde{\rho}(\varphi)$ and $\tilde{P}(\varphi)$ are from $\tilde{L}_m(\varphi)=\tilde{C}(\varphi)\tilde{L}^{(0)}_m(\varphi)$.

When $\tilde{U}(\varphi)\equiv 0$ and dividing by $\rho$, the Eq.~\eqref{eq22} becomes
\begin{align}\label{eq27}
\frac{2}{3-(\varphi')^2}\varphi''+(1-w)\varphi'=-(\alpha+\hat{\beta})(1-w),
\end{align}
where $w=\tilde{P}/\tilde{\rho}=P/\rho$.

Now, we consider the relation between the r.h.s. of the master equation \eqref{eq27} in Einstein frame and the first term of the r.h.s. of Eq.~\eqref{eq1} in Jordan frame in both cases, $\alpha+\hat{\beta}=0$ and $\alpha+\hat{\beta}\neq0$, respectively.
\\[6pt]
1) Case $\alpha+\hat{\beta}=0$
\\[6pt]
Taking into account the relations \eqref{eq3}, \eqref{eq9}, \eqref{eq12}, \eqref{eq23} and \eqref{eq24}, the quantity $\alpha+\hat{\beta}$ vanishes
\begin{align}\label{eq28}
\alpha+\hat{\beta}=\dfrac{d\ln A(\varphi)}{d\varphi}+\dfrac{d\ln C[\phi(\varphi)]}{d\varphi}=-\dfrac{1}{2}\left(\dfrac{d\ln F(\phi)}{d\phi}-2\dfrac{d\ln C(\phi)}{d\phi}\right)\dfrac{d\phi}{d\varphi}=0,
\end{align}
that is, the Eq.~\eqref{eq3} corresponding to the special class in Jordan frame is equivalent to $\alpha+\hat{\beta}=0$ in Einstein frame. Therefore, the master equation \eqref{eq27} becomes
\begin{align}\label{eq29}
\frac{2}{3-(\varphi')^2}\varphi''+(1-w)\varphi'=0.
\end{align}
\\[6pt]
2) Case $\alpha+\hat{\beta}\neq 0$
\\[6pt]
In previous subsection we found that vanishing  $\alpha+\hat{\beta}$ is equivalent to the special class of the scalar-tensor gravity in Jordan frame.

Consider the case where the condition \eqref{eq3} or \eqref{eq8} is not satisfied. To this end, find first the link between the coupling $\hat{\beta}$ and coupling parameter $\delta$ which was defined in Refs.~\cite{Guo2007, Majerotto2004, Amendola2007} and appeared in conservation equation
\begin{align}
\dot{\rho}(\phi)+3H[\rho(\phi)+P(\phi)]&=\delta H\rho(\phi), \\
\rho(\phi)&=\rho_0 a^{-3+\delta},
\end{align}
where $H$ is Hubble parameter, $\rho$ and $P$ are the matter density and pressure in Jordan frame(physical frame). The parameter $\delta$ which characterizes the scalar-matter direct coupling is constant. Ref.~\cite{Kim2017} has shown that the following relation holds
\begin{align}\label{eq32}
C(\phi)\sim a^6.
\end{align}

From \eqref{eq32}, we get
\begin{align}\label{eq33}
\hat{\beta}(\varphi)=\dfrac{d\ln C[\phi(\varphi)]}{d\varphi}=\dfrac{d\ln C}{d a}\frac{da}{d\varphi}=\dfrac{\delta H}{d\varphi/dt}.
\end{align}
Using the new evolution variable $p$, \eqref{eq21}, one obtains $\dfrac{d\varphi}{dt}=\varphi'\dfrac{dp}{dt}=\varphi'H$, and, therefore, we get
\begin{align}\label{eq34}
\hat{\beta}=\frac{\delta}{\phi'}.
\end{align}
In the cases that the condition Eqs.~\eqref{eq3} or \eqref{eq8} is not exactly satisfied, one can estimate $\hat{\beta}$ and $\alpha+\hat{\beta}$ if we are possible to determine the parameter $\delta$ from observation.

To this end, in the matter equation Eq.~\eqref{eq27}, assume the friction term to dominate as Ref.~\cite{Damour1993} has done, and suppose $w=0$. Then we arrive at a simple relation
\begin{align}\label{eq35}
\phi'=-(\alpha+\hat{\beta}).
\end{align}
Substituting relation \eqref{eq35} into \eqref{eq34}, we can estimate $\hat{\beta}$ provided that one is familiar with coupling strength $\alpha$ and coupling parameter $\delta$. From Eqs.~\eqref{eq34} and \eqref{eq35}, one obtains $\hat{\beta}=\frac{-\alpha\pm\sqrt{\alpha^2-4\delta}}{2}$. However, as $\hat{\beta}=0$ for $\delta=0$ from \eqref{eq34}, we take only $\hat{\beta}=\frac{-\alpha+\sqrt{\alpha^2-4\delta}}{2}$. Therefrom, we can obtain the following estimation for $\alpha+\hat{\beta}$ in the case of positive $\delta$  ,
\begin{align}\label{eq36}
\alpha+\hat{\beta}=\frac{\alpha+\sqrt{\alpha^2-4\delta}}{2}\leqslant\alpha, \qquad(\delta\geqslant0)
\end{align}
provided that an inequality $\alpha/4\geqslant\delta$ holds. The necessity of non-imaginarity of coupling $\hat{\beta}$ requires an inequality
\begin{align}\label{eq37}
\alpha/4\geqslant\delta,\qquad(\delta>0)
\end{align}

For the case where $\hat{\beta}$ is imaginary, the coupling $C(\phi)$ will be an oscillatory function of the scalar field. For the negative coupling parameter,
\begin{align}\label{eq38}
\alpha+\hat{\beta}=\frac{\alpha+\sqrt{\alpha^2+4\vert\delta\vert}}{2}>\alpha, \qquad(\delta<0)
\end{align}
Thus, in friction dominated approximation, we obtain two cases, $\alpha+\hat{\beta}\leqslant\alpha$ and $\alpha+\hat{\beta}>\alpha$.

Likewise, in general case where no approximation is made, the coupling parameter $\hat{\beta}(\varphi)$ also modifies the behavior of the force term represented by the r.h.s. of equation \eqref{eq27}, and the coupling parameter   $\hat{\beta}(\varphi)$ either increases or decreases the role of the only coupling strength $\alpha(\varphi)$. More important is that, at difference from the case of the only coupling $\alpha(\varphi)$, on account of the coupling $\hat{\beta}(\varphi)$, $\alpha(\varphi)+\hat{\beta}(\varphi)$ could change the point where $\alpha(\varphi)=da(\varphi)/d\varphi$ vanishes. In Ref.~\cite{Damour1993}, the authors had found that scalar-tensor theories of gravity contain generically a natural attractor mechanism tending to drive the gravity toward a state close to the general relativity. It is controlled by the following mechanism. 

Since the coupling strength $\alpha(\varphi)$ is the gradient of the potential  $a(\varphi)\equiv\ln A(\varphi)$, if there is a minimum of the potential  $a(\varphi)$, the scalar field would end up being caught at a minimum of $a(\varphi)$ where the gradient $\alpha(\varphi)=da(\varphi)/d\varphi$ vanishes. In our case, however, the addition of a new function $\hat{\beta}(\varphi)=d\ln C(\varphi)/d\varphi$, if the minimum of the function $d\ln C(\varphi)$ does not coincide with the one of the potential $\ln A$ or does not exist at all and it is a monotonic function of scalar $\varphi$, the attraction behavior toward the general relativity would change.

According to Ref.~\cite{Bertotti2003}, the current value of the coupling strength $\alpha_0$ is
\begin{align}
\alpha_0\leqslant\oldstylenums{1.4\times10^{-2}}
\end{align}
On the contrary, the estimation of the coupling parameter $\delta$ is not satisfactory yet. There are a few estimates for $\delta$(cf. Refs.~\cite{Kim2017,Guo2007}). Ref.~\cite{Guo2007} has given the value in an interval
\begin{align}
\oldstylenums{-0.08<\delta<0.03}\ (95\%\ C.L.)
\end{align}
for constant coupling models and
\begin{align}
\oldstylenums{-0.4<\delta<0.1}\ (95\%\ C.L.)
\end{align}
for varying coupling models ($\delta_0$ is a present value). The best-fit value is $\oldstylenums{\delta=-0.03}$ for the former and $\oldstylenums{\delta_0=-0.11}$ for the latter. According to our recent calculation, the constant coupling model has given
\begin{align}
\oldstylenums{\delta_0=0.038}\ (95\%\ C.L.)
\end{align}
Such uncertainties of the estimated coupling parameter $\delta$ does not make us do a definite conclusion as to the coupling $\hat{\beta}$ and $\alpha+\hat{\beta}$, even in the friction dominated approximation yet. 

\section{Conclusion}\label{sec4}

The above consideration shows that the existence of the scalar-matter direct coupling modifies the behavior of the coupling strength $\alpha(\varphi)$.

The scalar field as an ingredient of the gravitational field will be attracted toward a minimum of combined potential $\ln A(\varphi)+\ln C(\varphi)$ derivative of which yields $\alpha(\varphi)+\hat{\beta}(\varphi)$(see Eqs.~\eqref{eq23}, \eqref{eq24} and \eqref{eq28}).

The investigation on the evolutionary tendency of the coupling $\ln C(\varphi)$ is still at a beginning. But the fact that there is a lot of observational evidences on the existence of the scalar-matter direct coupling makes us attract an attention to the influence of the coupling $\hat{\beta}(\varphi)$ on the evolution of the scalar.

The magnitude of $\alpha(\varphi_0)$($\varphi_0$ is the present cosmological value of $\varphi$) measures the strength of the scalar ingredient in the gravitational interaction, i.e. the deviation from general relativity\cite{Damour1992}. As in friction dominated approximation of Eqs.~\eqref{eq36} and \eqref{eq38}, in 1general, the coupling $\hat{\beta}$ either reduces the effect of the only strength $\alpha$, $\alpha+\hat{\beta}<\alpha$ or increases the one, $\alpha+\hat{\beta}>\alpha$.

Consider a definite epoch, e.g. the present one. So small current value of the coupling strength as $\alpha_0=\oldstylenums{1.4\times10^{-2}}$ therefore may be explained also due to a combined effect of $\alpha$ and $\hat{\beta}$. For example, although the coupling strength $\alpha_0$ may be large as $\alpha_0=\oldstylenums{0.1}$, but negative $\hat{\beta}_0=\oldstylenums{-0.086}$ might give the current small value $\alpha_0+\beta_0=\oldstylenums{0.014}=\langle\alpha_0\rangle$ where brackets mean an ambiguity of sense.

In conclusion, our consideration substantiates the equivalence between Jordan frame and Einstein frame of the generalized scalar-tensor gravity with scalar-matter direct coupling in that the recovering Einstein tensor gravity due to the condition \eqref{eq3} or \eqref{eq8} is the same in either Jordan frame or Einstein frame as Eq.~\eqref{eq28} shows. 

A more accurate estimation of coupling parameter $\delta$ will give a realistic result for the evolution of scalar field.

\section*{Acknowledgement}

A part of this work had been performed on a visiting time of one of authors (K. Jik-su) to ICRA at Rome. He is very grateful to Prof. Ruffini for the hospitality during his stay.

\end{document}